\documentclass[prb,amsmath,superscriptaddress,twocolumn,showpacs]{revtex4-1}
\usepackage{bm}
\usepackage{pifont}
\usepackage{amssymb}
\usepackage{colordvi}
\usepackage{graphicx}
\usepackage{color}

\usepackage{color}
\definecolor{orange}{rgb}{1,0.5,0}
\definecolor{darkgreen}{rgb}{0,0.5,0}
\definecolor{darkblue}{rgb}{0,0,0.5}
\definecolor{purple}{rgb}{0.35,0,0.35}

\newcommand{\ov}[1]{\overline{{#1}}}
\newcommand{\be}{\begin{equation}}
\newcommand{\ee}{\end{equation}}
\newcommand{\bea}{\begin{eqnarray}}
\newcommand{\eea}{\end{eqnarray}}

\newcommand{\ave}[1]{\left\langle #1\right\rangle}
\newcommand{\ome}{\omega}

\def\nn{\nonumber}

\begin{document}


\title{Three-terminal semiconductor junction thermoelectric devices:\\
  improving performance}

\author{Jian-Hua Jiang}
\affiliation{Department of Condensed Matter Physics, Weizmann Institute of
  Science, Rehovot 76100, Israel}
  \affiliation{Department of Physics, University of Science and Technology of China, Hefei, Anhui 230026, China}
  \affiliation{Department of Physics, University of Toronto, 60 St. George St., Toronto, Canada ON M5S 1A7}
\author{Ora Entin-Wohlman}
\affiliation{
Raymond and Beverly Sackler School of Physics and Astronomy, Tel Aviv University, Tel Aviv 69978, Israel
}\affiliation{Department of Physics and the Ilse Katz Center for Meso-
  and Nano-Scale Science and Technology, Ben Gurion University, Beer
  Sheva 84105, Israel}
\author{Yoseph Imry}
\affiliation{Department of Condensed Matter Physics, Weizmann Institute of
  Science, Rehovot 76100, Israel}

\date{\today}

\begin{abstract}

A three-terminal thermoelectric device based on a $p$-$i$-$n$
semiconductor junction is proposed,   where the intrinsic region is
mounted onto a, typically bosonic,  thermal terminal.  
Remarkably, the
figure of merit of the device is governed also by the energy distribution
of the {\em bosons} participating in  the transport processes, in addition to the
electronic one.  
An enhanced figure of merit can be obtained when
the relevant distribution is narrow and the electron-boson coupling is strong
(such as for optical phonons). We study the conditions for which the figure of
merit 
of the three-terminal junction can be
greater than those of the  usual thermoelectric devices made of the
same material. A possible setup  with a  high figure of merit,
based on Bi$_2$Te$_3$/Si  superlattices, is proposed.

\end{abstract}

\maketitle

\vspace{-1cm}

 $^*$ patent application pending. 
 
 \vspace{.4cm}

\section{Introduction}
Thermoelectric energy conversion\cite{Honig,nonlinear} has stimulated for decades 
considerable research on fundamentals and applications. For a 
long time, people strove to find good thermoelectric materials with
high thermal to electric energy conversion efficiency. It has been found
that the optimal efficiency of a thermoelectric device in the
linear-response regime is\cite{Honig,nonlinear}
\be
\eta_{\rm opt} = \eta_C \frac{\sqrt{1+ZT}-1}{\sqrt{1+ZT}+1} \ ,\label{DEFF}
\ee
with $\eta_C$ being the Carnot efficiency. The optimal efficiency $\eta_{\rm opt}$ is an
increasing function of the figure of merit $ZT$. However,  $ZT=
T\sigma S^2/\left(\kappa_e+\kappa_{p}\right)$ is limited by
several competing transport coefficients, the conductivity
$\sigma$, the Seebeck coefficient $S$, and the electronic (phononic) thermal
conductivity $\kappa_e$ ($\kappa_p$), making high values of $ZT$
hard to achieve.\cite{nat.mat,dress} Mahan and Sofo (henceforth ``M-S'')
\cite{mahan} proposed to analyze and achieve high values of  $ZT$ by
separating it into two factors: (A) $T\sigma S^2/\kappa_e$ and (B)
$\kappa_e/(\kappa_e + \kappa_p )$. By recognizing that the electronic
transport quantities, $S$ and $\kappa_e/\sigma$, are related to the
mean and the variance of $E-\mu$ (i.e., the heat transferred by an
electron at energy $E$ with $\mu$ being the equlibrium value of the  chemical
potential\cite{ziman}), over the transport distribution function,
namely the energy-dependent conductivity $\sigma(E)$, they were able
to obtain\cite{mahan} 
\be
\frac{T\sigma S^2}{\kappa_e} = \frac{\ave{E-\mu}^2}{
  \ave{(E-\mu)^2}-\ave{E-\mu}^2} \ .\label{DEF}
\ee
For a quantity ${\cal O}$ (which is a function of $E$ )
$\ave{{\cal O}} = \int dE \sigma(E)(-\partial f_0/\partial E) {\cal O}(E)/\int
dE (-\partial f_0/\partial E) \sigma(E)$, 
with $f_0$ being the
equilibrium Fermi distribution function. According to M-S,\cite{mahan}
high $ZT$ values can be achieved (i) by increasing the factor (A)
through decreasing the variance of $E-\mu$ via a sharp structure in $\sigma(E)$,
away from the chemical potential $\mu$, and (ii) by reducing
the ratio $\kappa_p/\kappa_e$. Following this, there were many
proposals to achieve effectively narrow electronic bands, especially in
nanostructures with transmission resonances  and where the enhanced scattering of phonons at interfaces
also reduces the phononic heat
conductivity $\kappa_p$.\cite{Dresselhaus,interface,models,Linke} However, narrow
electronic bands do not necessarily lead to high $ZT$
values. Specifically, when $\kappa_e\ll \kappa_p$, $ZT$ does {\em not}
increase via reducing the variance of $E-\mu$ when it is already limited by $\kappa_p$, or if $\sigma$ is concurrently
decreased, it has been argued\cite{a-mahan} to even harm $ZT$  and
reduce the power factor $\sigma S^2$. The best figure of merit can
only be  obtained by considering the competition of all these
factors.\cite{a-mahan} 

One should be aware that the energy-dependent conductivity $\sigma(E)$ is 
well-defined  only  for {\em elastic} processes. In the direct generalization
of the M-S results to include the {\em inelastic} processes, $E$ in
Eq.~(\ref{DEF}) effectively becomes the average, $\ov{E}=(E_i+E_f)/2$ 
of the
initial and final energies ($E_i$ and $E_f$, respectively) of the
transferred electron.\cite{ZO} Nontrivial aspects of the {\em
  inelastic} processes are revealed in the ``three-terminal
thermoelectric devices'' proposed very
recently.\cite{3t0,3t1,3t2,3t,3t3,pre} By ``three-terminal'' we mean a 
set-up with an additional thermal terminal  supplying bosons (e.g.,
phonons, electron-hole excitations) involved in the inelastic processes, besides the two electronic
terminals. {In such devices, 
in addition to the normal thermoelectric effect
in the two electronic
terminals, 
there can be a thermoelectric effect due to the
energy transfer {\em between the thermal terminal and the
  electronic ones}. Physically, this is because the energy exchange between the
  electronic and bosonic systems induces an electric current, or vice
  versa. The optimal efficiency of such a three-terminal 
thermoelectric device in  the linear-response regime was found in
Ref.~\onlinecite{3t} to be the same as given by Eq.~(1) but with $ZT$
replaced by the three-terminal figure of merit $\tilde{Z}T$. By optimizing 
%
the efficiency of, say, a refrigerator working at equal temperatures of the two electronic reservoirs
it is found that the
%
%
three-terminal figure of merit is
\be
\tilde{Z}T=\frac{\ave{\ome}^2}
{\ave{\ome^2}(1+G_{el}/G_{in})(1+K_{pp}/K_{pe})-\ave{\ome}^2} \ .
\label{zt3t}
\ee
 The average in Eq. (\ref{zt3t})
is taken with respect to the conductance of each inelastic transport
channel. Specifically for a quantity as a function of the initial and final
energies ${\cal O}(E_i,E_f)$
\be 
\ave{{\cal O}}=\frac{\int dE_i dE_f g_{in}(E_i,E_f) {\cal O}(E_i,E_f)}{\int
 dE_i dE_f g_{in}(E_i,E_f)} \label{ave}
\ee
with 
$g_{in}(E_i,E_f)$ being the conductance of the
inelastic channel with given initial and final energies, and
$\ome=E_f-E_i$ is the energy
of the boson (also equal to the energy change of the carrier)  in 
each inelastic process .
This generalizes   
 the results of 
 Ref.~\onlinecite{3t} where only a
single inelastic transport channel has been considered.
In Eq. (\ref{zt3t})
$G_{el}$ ($G_{in}$) is the total
elastic (inelastic) conductance, $K_{pp}$ is the
purely boson-mediated thermal conductance  between the thermal
terminal and the other two terminals, and
$K_{pe}=e^{-2}G_{in}\ave{\ome^2}$ is the thermal  
conductance characterizing the  heat transfer between the bosons and the
electrons. 
This is the generalization of the theory of
M-S to the three-terminal case where the principal quantity is now the
{\em energy change} $\ome$. A direct consequence is that there is
{\em no} cancelation of the electron and hole contributions to the
three-terminal thermopower. We find that a high three-terminal figure
of merit $\tilde{Z}T$ requires: (i) the dominance of the inelastic
transport $G_{in}\gg G_{el}$, (ii) a small variance of the energy change
$\ave{\ome^2}-\ave{\ome}^2\ll \ave{\ome}^2$, (iii) a large  ratio of
$K_{pe}/K_{pp}$ which can be realized when $G_{in}$ and $\ave{\ome^2}$
are large or $K_{pp}$ is small. Small $K_{pp}$ values should be
achievable by, e.g., engineering the interfaces between the central system and the
two electronic terminals. Remarkably, the purely electronic heat conductance, $K_e$
does not appear and does not need to be small!

In addition  to the pursuit of
a narrow distribution of $\sigma(E)$ in the M-S proposal, the three-terminal
figure of merit may also benefit from a ``selection'' of the energy change
$\ome$ either via the electronic structure or via the bosonic spectrum
so that the variance of $\ome$ can be small. This can be
achieved also by a small bandwidth of the bosons involved in the
inelastic transport. The
merits of the three-terminal configuration are several. (i) There
may be {\em no} restriction on the effective electronic bandwidth [or
other parameters required for a small variance of $\sigma(E)$] as the
electronic thermal conductivity $\kappa_e$ does {\em not} appear in
the three-terminal figure of merit; (ii) Smaller effective boson
bandwidths usually make the bosonic thermal conductance 
$K_{pp}$ smaller, which {\em improves} $\tilde{Z}T$; (iii) In general,
if, e.g., due to  momentum or energy conservation (a ``selection''),
only the bosons in a small energy range are involved in the
transitions, the effective bandwidth can also be small. As a possible
candidate, optical phonons have small bandwidths (see Ref.~\onlinecite{LB}) and
their coupling with carriers is relatively strong. For the p-n configurations discussed here, this necessitates an electronic band gap smaller or of the order of the phonon frequencies. This can happen {\it e.g.}  in solid solutions of the HgCdTe family. \cite{LB}  
Further examples, such as superlattices, are mentioned below. 
For acoustic phonons the
coupling to the carriers is usually stronger at large
wavevectors/frequencies\cite{LA} (e.g., around the Debye frequency)
where the density of states of phonons is also large.  In Refs.~\onlinecite{3t2,3t3}, the Coulomb
interaction  between the quantum-dot system and the lead(s)  plays the same role as bosons to
induce the inelastic transport.


The choice of the thermal terminal is very important. It should provide bosonic excitations with an energy matching the required electronic one, $E_g$. When $k_BT\ll E_g$, and for upgoing transitions the product of the effective $E_g$ and the exponentially small boson population has to be optimized (as done for example in  Ref. \onlinecite{mahan}). In addition, the Debye energy is often not high enough, and multiple phonon processes  are quite weak. This can be remedied by using optical phonons or by using a material with $\ome_D\gtrsim E_g$
with $\ome_D$ being the Debye energy (e.g., diamond, whose
$\ome_D=190$~meV) for the thermal terminal. 
An electronic thermal bath where its electron-hole pair excitations or
plasmons interact with the electrons in the intrinsic region,
is another possibility.\cite{3t2} The direct tunneling thermal
conductance between the thermal terminal and the {other} two terminals
can  be made exponentially small via controlling the
geometry of the contact. Here we suggest and advocate optical phonons. 
We will mainly consider the
optical phonon bath without implying  that it is the only possibility.

We thus propose a three-terminal device based on the $p$-$i$-$n$ junction
where the intrinsic region is contacted with, {\it e.g.,} a phonon source or a thermal terminal, as above. 
In Sec.~\ref{SEC3T}, we present the
device structure and show how the proposal works for semiconductors
and some of their superlattices where the band gap may be smaller than the phonon
energy. We estimate the figure of merit 
and find
that such a device can have a better performance than the usual
two-terminal device made of the {\em same} material. We conclude this
study in Sec.~\ref{SECCON}. The discussions throughout this paper
are focused on the linear-response regime. Nevertheless, if the system
is not far away from that regime, such treatment  may still offer useful
information.\cite{nonlinear}

\section{Three-terminal $p$-$i$-$n$ junction thermoelectric device}

\label{SEC3T}

\subsection{The device concept and structure}

In the following discussions we specifically consider a
$p$-$i$-$n$ junction made of ``extremely narrow-gap
semiconductors''. The structure is depicted in Fig.~\ref{FIG1}a, for the 
case of converting thermal to electrical energy. It
can be viewed as an analog of the $p$-$i$-$n$ photodiode, where
photons are replaced by phonons, {\it i.e.}, the phonon-assisted inter-band
transitions lead to current generation in the junction. The device can
also be used to {\it  e.g.} cool the thermal terminal via the electric
  current between the electronic terminals. We
focus on the situation where single-phonon assisted inter-band
transitions are allowed. (The electronic band gap is hence required to
be {\em smaller} than the phonon energy). Such processes may play 
a significant role in the transport across the junction whereas the
transition rates of multi-phonon processes are much smaller. In
semiconductors, the optical phonon energy is usually in the range of $20\sim 
100$~meV. There are several 
candidates with such small band gaps:
(i) Gapless semiconductors resulting from accidental band degeneracy
in solid solutions, such as Pb$_{1-x}$Sn$_x$Te and
Pb$_{1-x}$Sn$_x$Se.\cite{gaplesssemi} At a certain    mole fraction $x$ the
band gap closes, around which 
it can be very small. (ii) Gapless
semiconductors originating from band inversion, such as HgTe and
HgSe.\cite{gaplesssemi} The  band gap can be tuned via the quantum-well
effect in the superlattices composed of a gapless semiconductor and
a normal semiconductor without band inversion.\cite{bhz} An example is
the HgTe/CdTe superlattices with  a tunable band gap.\cite{balents} (iii)
 Multilayers (superlattices)  of ``topological" insulators with 
ordinary insulator layers sandwiched between the topological
ones.\cite{burkov} For example, in Bi$_2$Te$_3$/Si superlattices the
energy gap can be tuned by varying the 
thickness of the topological and ordinary layers.\cite{burkov}
A significant merit of superlattices is that the lattice thermal
conductivity along the growth direction can be much smaller than both
of the two bulk constituents. For example, the Si/Ge superlattices have a
thermal conductivity about two orders of magnitude smaller than the
bulk values.\cite{lowkl} We also point out that the same idea 
can be applied to devices where the role of optical phonons is
played by other bosons. If the energy of such bosons is higher the
requirement for small band gap can be softened.

For simplicity we consider a linear junction where the conduction
band edge varies linearly in the intrinsic region with the coordinate
along the junction $z$ from $z=-L/2$ to $L/2$ as 
\be
E_c(z) = E_g(1/2-a z/L) \label{ecza}
\ee
with $E_g$ being the band gap.\cite{Sze}    We set
$0<a<1$ so that the $p$-doped ($n$-doped) region is at the left
(right) side of the junction [see Fig.~\ref{FIG1}b]. These electronic
terminals can have temperatures different from that of the thermal
terminal. It is favorable to bend the two electronic terminals away
from the thermal one so that they will be better thermally isolated
from the thermal terminal and from each other [see Fig.~\ref{FIG1}a]. 
Another possibility is a ``thermal finger" for the boson bath [Fig.~\ref{FIG1}c], which can 
be well isolated from the electronic leads.

\begin{figure}[htb]
\hspace{-0.4cm}\includegraphics[height=3.4cm]{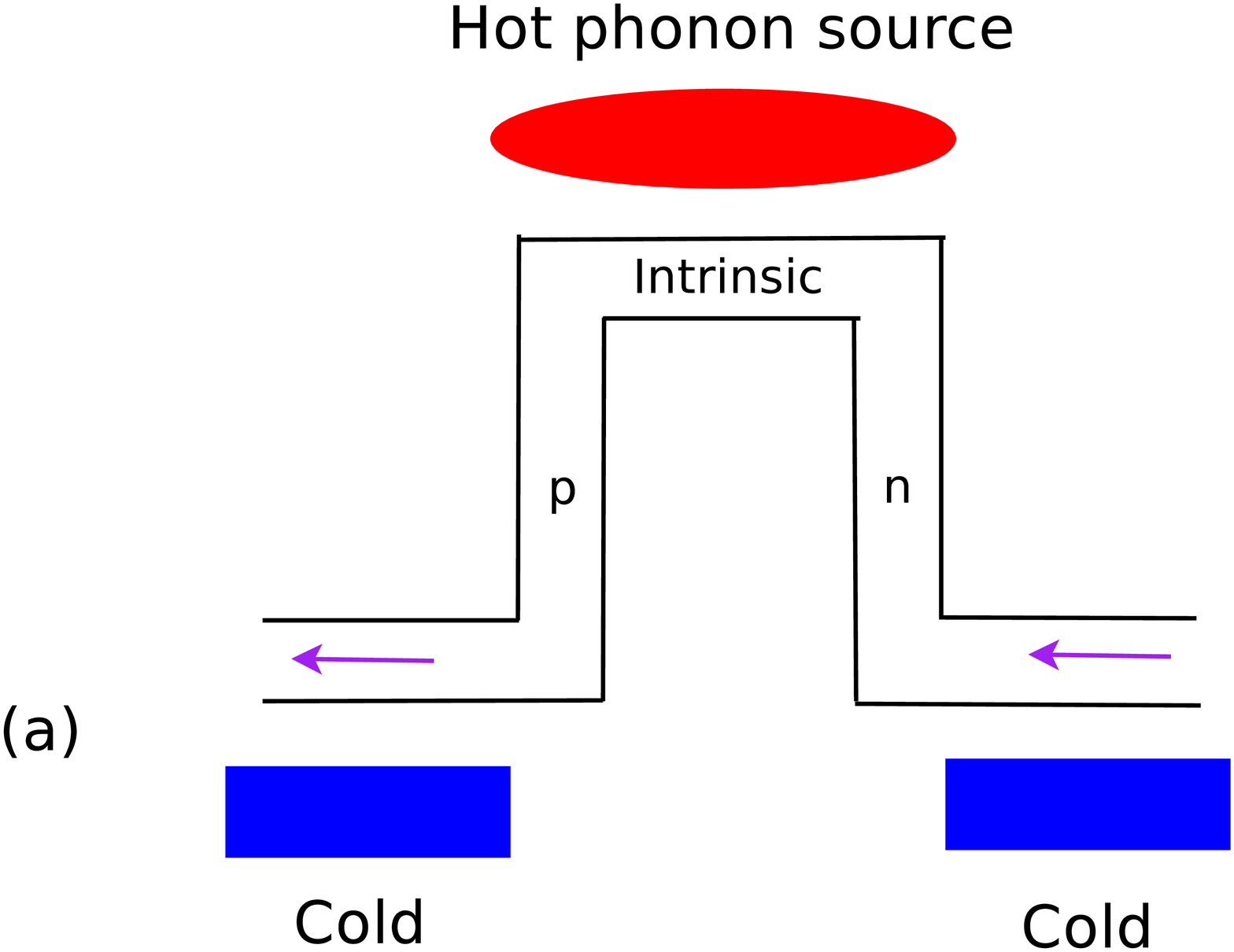}
\vskip 0.4cm
\hspace{-0.42cm}\includegraphics[height=1.8cm]{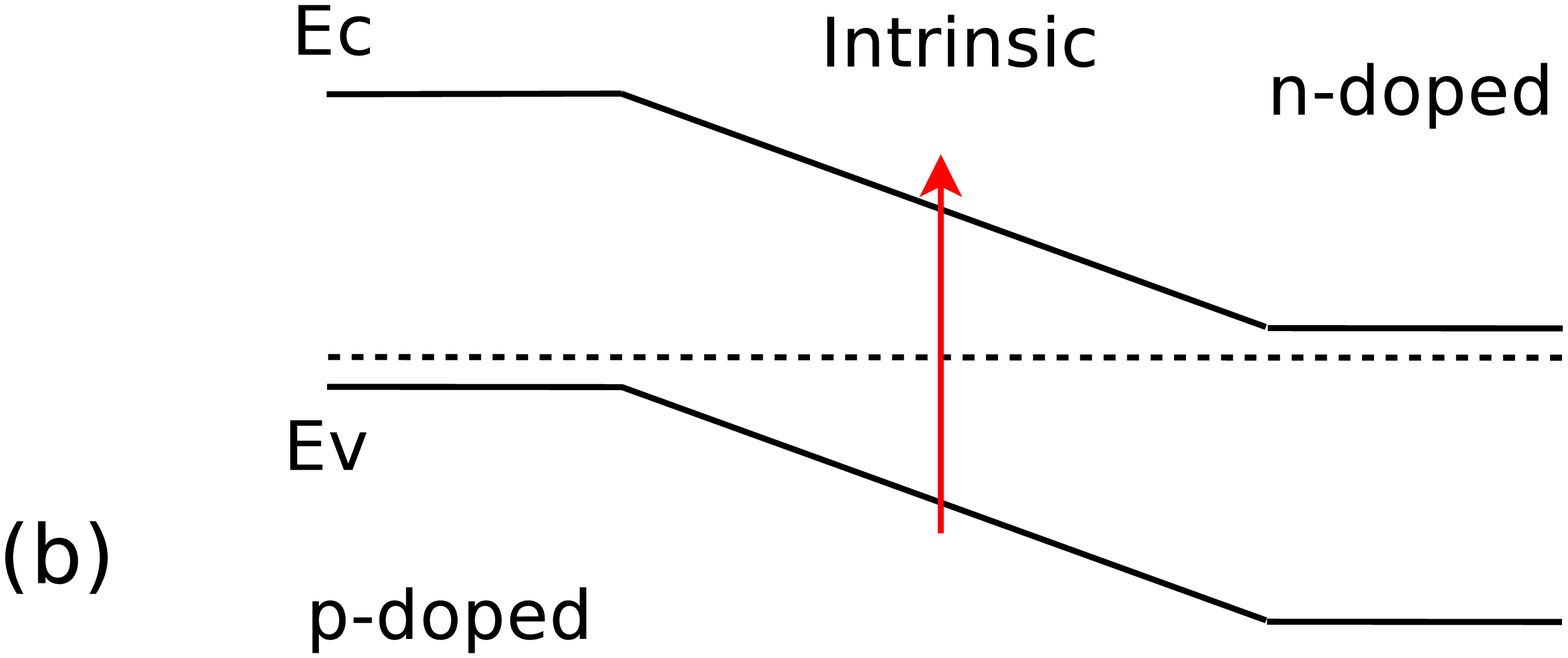}
\vskip 0.5cm
\hspace{-0.5cm}\includegraphics[height=1.5cm]{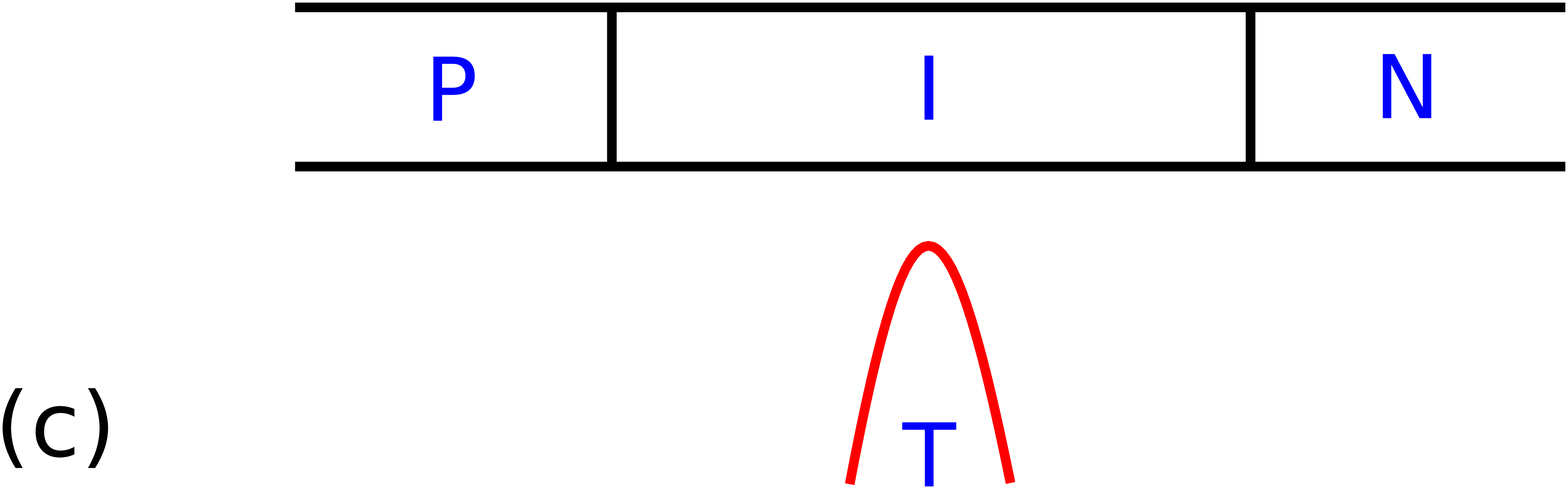}
\caption{(Color online) (a) A schematic illustration of a possible
  three-terminal $p$-$i$-$n$ junction thermoelectric device. The
  phonon source acts as the thermal terminal. The two electronic
  terminals are the $p$ and $n$ doped regions, respectively. As an
  example we illustrate the situation where the device converts thermal
  energy from the thermal terminal to electrical energy. The arrows denote the
  direction of the electric current. (The device
  can also cool the thermal terminal by consuming the electrical
  energy, not directly indicated in the figure).  (b) The band structure of the
  $p$-$i$-$n$ junction. The dotted line is the chemical potential at
  equilibrium.  The red arrow labels the phonon-assisted inter-band
  transition which generates electrons and holes. When drifting with the
  built-in electric field the generated nonequilibrium carriers lead
  to current flow across the junction.
  (c) A schematic illustration of another possible set-up of
  three-terminal $p$-$i$-$n$ junction thermoelectric device. The
  thermal terminal (labeled as ``T'' in the figure) is a thermal
  finger, which can be either a phonon reservoir or an electronic
  one. In the latter case the electron-hole pair excitations or plasmons are
  coupled with the electrons in the intrinsic region via the Coulomb
  interaction.
} \label{FIG1}
\end{figure}

\subsection{``Ideal" figure of merit}
We start by considering only the phonon-assisted transport, ignoring
the phononic thermal {conduction} and the normal diode transport. The
phonon-assisted inter-band transitions generate current flow in the
junction.   In the
linear-response regime, the thermoelectric transport equations are
written as\cite{3t}
\be
  \left( \begin{array}{c}
      I_e\\ I_{Q}^e\\ I_{Q}^{pe}\end{array}\right) =
  \left( \begin{array}{cccc}
      G_{in} & L_1 & L_2 \\
      L_1 & K_{e}^0 & L_3 \\
      L_2 & L_3 & K_{pe} \\
    \end{array}\right) \left(\begin{array}{c}
      \delta\mu/e \\ \delta T/T\\ \Delta T/T \end{array}\right)\  .
\ee
Here $I$ and $I_{Q}^e$ are the electronic charge and heat currents
flowing between the two 
electronic terminals,  $e<0$ is the electronic
charge,  $I_{Q}^{pe}$ is the heat current from the thermal terminal to
the two 
electronic ones,  $G_{in}$ is the conductance in the inelastic channels, $K_e^0$ is related to
the electronic heat conductance between the 
electronic terminals, and
$K_{pe}$ is that between the thermal terminal and the 
electronic ones. The transport coefficients 
$L_1$, $L_2$, and $L_3$ are related to the currents induced by the temperature differences
(thermopower effect) and the  current-induced temperature 
differences (refrigerator and heater effects). {$\delta\mu=\mu_L-\mu_R$ 
($\delta T=T_L-T_R$) is the chemical potential (temperature)
difference between the two 
electronic terminals, and $\Delta T=T_p-\frac{1}{2}(T_L+T_R)$ is the difference between the
temperature of the thermal terminal and the average temperature of the two
electric ones}, with $T_L$, $T_R$, and $T_p$ being the temperatures of
the left, right electronic terminals, and the phonon terminal, 
respectively. With these definitions of $\Delta T$, $\delta T$, and $I_{Q}^e$,
$I_Q^{pe}$ the Onsager reciprocal relationships  are satisfied. 
In such a
setup, as found in Ref.~\onlinecite{3t}, the three-terminal Seebeck
coefficient and figure of merit are, when $G_{el}$ and
  $K_{pp}$ are neglected,
\bea
S_p = \frac{L_2}{TG_{in}}\ , \quad \quad \tilde{Z}T = \frac{
  L_2^2}{G_{in}K_{pe} - L_2^2} \ ,
\label{ideal}
\eea
respectively. To obtain the transport coefficients and the figure of
merit, we need to calculate the phonon-assisted currents through the
system.

The Hamiltonian of the inter-band electron-phonon coupling 
is\cite{more}
\be
H_{e-ph} = \frac{1}{\sqrt{V}}\sum_{{\bf k}{\bf q}\lambda \nu \rho} M_{{\bf
    q}\lambda \nu \rho}c^\dagger_{{\bf k+q}\rho}c_{{\bf k}\nu}(a_{{\bf q}\lambda}
+ a^\dagger_{-{\bf q}\lambda}) + {\rm H.c.}\ ,
\label{hami}
\ee
where $V$ is the volume of the system, $\lambda$ is the phonon branch
index, $\nu$ ($\rho$) runs through the valence (conduction) band indices, $M_{{\bf q}\lambda \nu\rho }$ is the matrix element of the electron-phonon
coupling, and $c^\dagger$ ($a^\dagger$) is the electron (phonon)
creation operator. Due to momentum (when valid) and energy conservations, phonons
involved in such processes will be in a small energy range. For
indirect-band semiconductors and with momentum conservation, 
these phonons can be  acoustic as well
as optical ones. For simplicity and   definiteness, we consider a
direct-band semiconductor system and assume that the contribution from
the optical phonons is the dominant one.
From Eq.~(\ref{hami}), the net electron-hole generation rate per unit
volume, $g_p$, is given for single-phonon transitions by the
Fermi golden rule as
\bea
g_p &=& \frac{2\pi}{\hbar} \sum_{{\bf k}{\bf q}\lambda \nu \rho}
|M_{{\bf
    q}\lambda \nu \rho}|^2 \Big[(1-f_{{\bf k+q}\rho})f_{{\bf k}\nu}N_{{\bf q}\lambda} 
 - f_{{\bf k+q}\rho}
\nn\\ && \mbox{}
\times (1-f_{{\bf k}\nu})(N_{{\bf q}\lambda} +1) \Big]\delta(E_{{\bf k+q}\rho}-E_{{\bf
    k}\nu}-\ome_{{\bf q}\lambda})\nn\\
&=& \int dE_i dE_j \Gamma_{p}(E_i,E_j) \Big[(1-f(E_j))f(E_i)N(\ome_{ji})
\nn\\ && \mbox{}
-f(E_j)(1-f(E_i))(N(\ome_{ji})+1)\Big] ,
\label{gp1}
\eea
where
\bea
\Gamma_{p}(E_i,E_j) &=& \frac{2\pi}{\hbar} \sum_{{\bf k}{\bf q}\lambda \nu \rho}
|M_{{\bf q}\lambda \nu \rho}|^2 \delta(E_j-E_i-\ome_{{\bf
    q}\lambda})\nn\\ &&\mbox{} \times \delta(E_{{\bf 
    k+q}\rho}-E_j) \delta(E_{{\bf k}\nu}-E_i) .
\eea
Here, $E_{{\bf k+q}\rho}$, $E_{{\bf k}\nu}$, and $\ome_{{\bf q}\lambda}$ are the electron
and phonon energies when the two systems are uncoupled,  $f$ and $N$ are the nonequilibrium
distributions of electrons and phonons in the intrinsic region, with 
$\ome_{ji}\equiv E_j-E_i>0$ due to energy conservation. According to
Ref.~\onlinecite{shockley}, when the length of the intrinsic region
$L$ is sufficiently smaller than the carrier diffusion length, the
electronic distribution in the conduction (valence) band in the
intrinsic region can be well approximated as the distribution in the
$n$-doped ($p$-doped) electronic terminal.\cite{shockley} Similarly
the phonon distribution is almost the same as that in the thermal
terminal when the contact between the intrinsic region and the thermal
terminal is good. Finally, a small amount of 
disorder which always exists in real systems and relaxes the momentum
conservation can enhance the phonon-assisted inter-band transitions.

\par The transport coefficients are determined by studying the currents at
a given bias and/or a temperature difference. The key relation is the
continuity equation,\cite{shockley,closed}
\be
\partial_t n_\alpha(z,t) =
\frac{n_\alpha^{eq}(z)-n_\alpha(z,t)}{\tau_\alpha} -
\frac{1}{q_\alpha}\partial_z I_\alpha(z,t) + g_p\  ,
\label{cont}
\ee
where $n_\alpha$ ($\alpha=e,h$) are the electron and the hole densities
in the conduction and valence bands, and $n_\alpha^{eq}$ are the
equilibrium values of those densities. $I_\alpha$ are the charge
currents, $q_{\alpha}$ are the charges of the electron and the hole,
$\tau_\alpha$ are the carrier lifetimes limited by the recombination
processes other than the phonon-assisted ones that have already been
taken into account in $g_p$, and $g_p$ is the net carrier density
generation rate given in Eq.~(\ref{gp1}). The currents $I_{\alpha}$,
which consist of diffusion and drift parts, are
\be
I_\alpha(z) = -e \chi_\alpha n_\alpha(z) {\cal E} - q_\alpha
D_\alpha\partial_z n_\alpha(z) \ ,
\label{Ial}
\ee
where $\chi_\alpha$ and $D_\alpha$ are the mobilities and the diffusion
constants, respectively. They are related by the Einstein
relation, $D_\alpha=-(k_BT/e)\chi_\alpha$. ${\cal E}= aE_g/(eL)$ is
the built-in electric field in the intrinsic region.

If Boltzmann statistics for the electrons can be assumed everywhere, the net generation rate
$g_p$ will depend on $z$ very weakly such that its spatial
dependence can be ignored. In this situation, the total carrier densities 
can be divided into two parts, $n_\alpha=n_{\alpha,g}+n_{\alpha,n}$ where
$n_{\alpha,g}=g_p\tau_\alpha$ are the spatially-independent carrier
densities generated by the phonon-assisted inter-band transitions and
$n_{\alpha,n}$ are the ``normal'' densities in the 
junction,  
determined by the continuity equation with
$g_p=0$. Similarly, the current is divided into two parts,
$I_{\alpha}=I_n^\alpha+I_g^\alpha$. The current in the normal diode
channel can be obtained from Eqs.~(\ref{cont}) and (\ref{Ial}) with
proper boundary conditions, yielding the celebrated rectification
current-voltage relation, 
\be
I_n^\alpha = I_{ns}^{\alpha}(e^{-\delta
  \mu/k_BT}-1) \label{rec-iv}
\ee
with $I_{ns}^{\alpha}= - e D_\alpha L_\alpha^{-1}
n_\alpha^m $ being the saturated currents. Here $n_\alpha^m$ is the
density of the minority carrier and $L_{\alpha}$ is its diffusion
length.\cite{shockley} 

Inserting $n_{\alpha,g}$ into
Eq.~(\ref{Ial}), one obtains the currents in the phonon-assisted
channel\cite{note11}
\be
I_{g}^\alpha = - e \chi_\alpha g_p\tau_\alpha {\cal E} = - \chi_\alpha
\tau_\alpha E_gaL^{-1} g_p \  .
\ee
In the linear-response regime,
{
\bea
g_p &=& \int{dE_i}dE_j 
\Gamma_p(E_i,E_j) f^0(E_i)(1-f^0(E_j))
 \nn\\ && \mbox{} \times N^0(\ome_{ji}) 
 \Bigl[\frac{\delta \mu}{k_BT}+
\frac{\ov{E}_{ij} - \mu}{k_BT}\frac{\delta T}{T} +
\frac{\ome_{ji}}{k_BT}\frac{\Delta T}{T} \Bigr] \nn\\
&\equiv& g_p^0 \Bigl[\frac{\delta \mu}{k_BT}+
\frac{\ave{\ov{E}_{ij} - \mu}}{k_BT}\frac{\delta T}{T} +
\frac{\ave{\ome_{ji}}}{k_BT}\frac{\Delta T}{T} \Bigr] \ ,
\label{gpp}
\eea
where $f^0$} and $N^0$ are the equilibrium distribution functions of the
electrons and the phonons, respectively, 
and $\ov{E}_{ij}\equiv (E_i+E_j)/2$. Consequently 
\begin{align}
& G_{in} = -\frac{eE_g g_p^0 a}{k_BTL}\sum_\alpha \chi_\alpha
\tau_\alpha \ ,\quad 
L_1 = \frac{G_{in}}{e} \ave{\ov{E}_{ij}-\mu} ,\nn \\
& L_2 = \frac{G_{in}}{e} \ave{\ome_{ji}} ,\quad
K_e^0 =  \frac{G_{in}}{e^2} \ave{(\ov{E}_{ij}-\mu)^2} ,\nn \\
& L_3 = \frac{G_{in}}{e^2} \ave{ (\ov{E}_{ij}-\mu)\ome_{ji}} ,\quad 
K_{pe} = \frac{G_{in}}{e^2} \ave{\ome_{ji}^2} , 
\label{GLK}
\end{align}
where $g_p^0$ is the equilibrium transition rate [defined in
Eq.~(\ref{gpp})] and the average is defined in Eq.~(\ref{ave}) with
{\bea
g_{in}(E_i,E_j)&=&-\frac{eE_g
    a}{k_BTL} \Gamma_p(E_i,E_j) 
f^0(E_i)(1-f^0(E_j)) \nn\\
&& \mbox{}\times  N^0(\ome_{ji})\sum_\alpha \chi_\alpha
\tau_\alpha .\label{F1}
\eea}
The above results are very similar to those obtained in
Ref.~\onlinecite{3t}: due to the inelastic nature of the transport,
the carrier energies at the $p$ and $n$ terminals are different;
the heat transferred between the two terminals is the average one
$\ov{E}_{ij} - \mu$,  whereas the energy difference $\ome_{ji}$ is
transferred from the thermal terminal to the two electronic ones. This
is also manifested in the way  the temperature differences are coupled
to the heat flows\cite{3t} in Eq.~(\ref{gpp}), ensuring the Onsager
relations. An important feature is that the three-terminal Seebeck
coefficient, $L_2/(TG_{in})$, is {\em negative definite} since
$\ome_{ji}>0$. In
contrast, the two-terminal Seebeck coefficient, $L_1/(TG_{in})$, does not possess 
this property. It  can be positive or negative due to the
partial {\em cancelation} of the contributions from electrons
and holes, whereas there is no such cancelation {for} $L_2$ [see
Eq.~(\ref{GLK})]. Equation (\ref{GLK}) is a generalization of the
results in Ref.~\onlinecite{3t} where there was only a single
microscopic energy channel. When many inelastic processes
coexist, the contribution of each process is weighed by its 
conductance. From Eqs.~(\ref{ideal}) and (\ref{GLK}), we find that the
``ideal" three-terminal figure of merit is
\be
\left. \tilde{Z}T\right|_{\rm ideal} = 
\frac{\ave{\ome_{ji}}^2}{\ave{\ome_{ji}^2} - \ave{\ome_{ji}}^2 }\  .
\label{zt1}
\ee
For a single microscopic energy channel system where $\ome_{ji}$ is
fixed, this figure of merit goes to infinity.\cite{3t} When there
are many {many such energy channels}, it becomes finite due to the
nonzero variance {of $\ome_{ji}$}. We
estimate the figure of merit when $\langle\omega\rangle -E_g\simeq k_BT$ and $\gamma,
k_BT\ll \langle\omega\rangle $. {Here} $\langle\omega\rangle$ is
the average phonon energy and $\gamma$ {is the effective
  bandwidth (i.e., the variance 
of $\ome_{ji}$ due to spectral dispersion) of the involved phonons}.
{From Eqs.~(\ref{GLK}) and (\ref{F1}) one finds that the variance
  of $\ome_{ji}$ is limited by $\gamma^2$ or $(k_BT)^2$ whichever is
  smaller. For example, when the effective phonon bandwidth $\gamma$ is much
  smaller than $k_BT$, the variance is rather limited by $\gamma^2$.  
  In the situations where $\gamma, k_BT\ll \ave{\ome_{ji}}$, the
  numerator in Eq.(18) is much larger than the denominator.}
The ``ideal" figure of merit can be very high thanks to the
electronic band gap when $E_g\gg k_BT$ or the narrow bandwidth of the
optical phonons $\langle\omega\rangle\gg \gamma$. However, in realistic situations, as often happens,
the parasitic heat conduction is another major obstacle to a high figure
of merit. This   will be analyzed in the next subsection. 


\subsection{Realistic figure of merit}
Besides the phonon-assisted transport channel, there is the normal
diode channel which is dominated by  the (elastic) barrier transmission and the
diffusion of minority carriers. It contributes to $G$, $L_1$, and 
$K_e^0$ as well. In addition there are the ``parasitic'' heat currents
carried by phonons flowing between the two 
electronic terminals and
those from the thermal terminal to the electronic ones. Taking into account
all these, the thermoelectric transport equations are written as 
\begin{widetext}
\be
\left( \begin{array}{c}
    I_e\\ I_{Q}\\ I_{Q}^{T}\end{array}\right) =
\left( \begin{array}{cccc}
    G_{in} + G_{el}  & L_1 + L_{1,el} & L_2 \\
    L_1 + L_{1,el} & K_e^0 + K_{e,el}^0 + K_p & L_3 \\
    L_2 & L_3 & K_{pe} + K_{pp} \\
  \end{array}\right) \left(\begin{array}{c}
    \delta\mu/e \\ \delta T/T\\ \Delta T/T \end{array}\right)\  .
\ee
\label{tran3}
\end{widetext}
Here, $I_Q=I_Q^e+I_Q^p$ is the total heat current between the two 
electronic
terminals which consists of the electronic $I_Q^e$ and the phononic
$I_Q^p$ contributions; $I_Q^{T}=I_Q^{pe}+I_Q^{pp}$ is the total heat
current flowing out of the thermal terminal to the two electronic
ones with $I_Q^{pp}$ being the purely phononic part. 
Finally, $G_{el}$, $L_{1,el}$,
and $K_{e,el}^0$ are the contributions to the transport coefficients
from the normal diode (elastic) channel, and  $K_p$ and $K_{pp}$
are the heat conductances 
of phonons flowing between the two 
electronic
terminals and those flowing from the thermal terminal to the two
electronic ones, respectively.   Note that the elastic channel
does {\em not} contribute to $L_2$ and $L_3$ which 
are solely related to the inelastic processes. 
The three-terminal figure of merit can be obtained
by optimizing the efficiency of a refrigerator working at
$\delta T=0$ but with  finite $\delta\mu$ and $\Delta T$.  
This  figure of merit    is\cite{3t}
\be
\tilde{Z}T = \frac{L_2^2}{( G_{in} + G_{el} )( K_{pe} + K_{pp} ) - L_2^2 } \ , \label{TILZ}
\ee
which 
is equivalent to Eq.~(\ref{zt3t}). The
diode conductance is $G_{el} = -(e/k_BT)\sum_{\alpha} I_{ns}^{\alpha}$.
A high figure of merit requires $G_{in}\gg G_{el}$
which is not difficult to achieve according to the analysis
  in the next subsection (\ref{ESTIM}). In such a situation and when
$\gamma\ll \langle\omega\rangle$ or $k_BT\ll\langle \omega\rangle$, 
i.e., when the energy width due to
$k_BT$ or $\gamma$ gives a much weaker limitation to $\tilde{Z}T$ than
$K_{pp}$, one finds from Eq.~(\ref{TILZ})\cite{fotno1}
\be
\tilde{Z}T \simeq \frac{K_{pe}}{K_{pp}}\  .\label{f0}
\ee
Estimations carried out in Sec.~\ref{ESTIM} indicate
that there are parameter regimes where the figure of merit
Eq.~(\ref{TILZ}) can be greater than the
usual two-terminal ones in the {\em same} material. 
We repeat that, unlike the two-terminal case, the thermal conductance $K_p$ between the electronic terminals, does not affect the three-terminal figure of merit. 

Often $K_p$ and $K_{pp}$ 
are of the same order of
magnitude. We note that $K_p$ and $K_{pp}$ are small in several
gapless semiconductors such as PbSnTe, PbSnSe, BiSb, HgTe, and
HgCdTe. In addition, superlattice structures (and other planar
composite structures) usually have much lower $K_p$ and $K_{pp}$ along the growth 
direction than the bulk materials.\cite{spMahan} The
geometry where the electric current flows along that direction is
promising for high figures of merit.

\subsection{Estimation of the figure of merit}

\label{ESTIM}
We defined $\tau_\alpha$ ($\alpha=e,h$) as the carrier lifetimes due
to recombination processes other than the phonon-assisted ones that
have already been taken into account in $g_p$ [see
Eq.~(\ref{gp1})]. The carrier lifetimes due to the phonon-assisted
processes, $\tau_{e,p}$ for the electrons and $\tau_{h,p}$ for the holes,
satisfy the detailed balance relations
$g_p^0\tau_{e,p}=g_p^0\tau_{h,p}=n_i$ in the intrinsic region. Here
$n_i$ is the electron (hole) density in that  region and
$g_p^0$ is the equilibrium transition rate per unit volume defined in
Eq.~(\ref{gpp}). The transition between minibands in semiconductor
superlattices is usually dominated by the phonon-assisted processes in 
the dark limit when the miniband gap is smaller than the phonon
energy.\cite{ridley} We introduce the parameter $\zeta$ to write
$g_p^0\tau_e\simeq g_p^0\tau_h= \zeta n_i$.  This parameter is governed by
the electron-phonon interaction strength, being of order unity when
such a coupling is strong as is the case for some III-V (and other)
semiconductors, or smaller (it will be taken  as $1/4$
below).\cite{ridley} Inserting this into Eq.~(\ref{GLK}) one finds
\be
G_{in} \simeq e^2 (k_BT)^{-2} E_g \zeta n_i \sum_\alpha D_\alpha L^{-1} \  ,\label{ginap}
\ee
where 
$L$ stands for the length of the intrinsic
region which will be taken to be much smaller than the diffusion
length $L_\alpha$. \cite{shockley} In Eq. (\ref{ginap}) we  have chosen $a$ [defined in Eq.(\ref{ecza})] to
be close to 1 
which amounts to 
high doping density
$N_d\lesssim N_0\equiv n_i\exp[E_g/(2k_BT)]$ 
%
%
in the $p$ and $n$
doped regions. The elastic conductance is the slope of the $I$-$V$
rectification characteristics, Eq.~(\ref{rec-iv}), at $V=0$, 
\be
G_{el} \simeq e^2 (k_BT)^{-1} n_i^{2} N_d^{-1} \sum_\alpha D_\alpha
L_\alpha^{-1} \  ,  \label{gelap}
\ee
with $L_\alpha$ denoting the carrier diffusion length. 
The
majority carrier densities (doping densities) in both the $n$ and $p$
regions are taken to be $N_d$. It follows then that
\be
\frac{G_{el}}{G_{in}} \simeq \frac{k_BT}{E_g}\frac{n_i}{\zeta
  N_d}\frac{\sum_\alpha D_\alpha  L_\alpha^{-1}}{\sum_\alpha D_\alpha
  L^{-1} } \ .
\ee
A small ratio can be easily achieved since $L < L_{\alpha}, k_BT<E_g$ and
$n_i\ll N_d$. Therefore the contribution of $G_{el}$ is not the main
obstacle for a high figure of merit in this device.

For $G_{el}\ll G_{in}$ with $\gamma$ (the effective width of the relevant energy
band) and $k_BT$ being much smaller than $\langle\omega\rangle$ (the typical
phonon energy), the figure of merit $\tilde{Z}T$  is given by Eq.~(\ref{f0}).
The phononic heat conductance $K_{pp}$ is, as usual, the least known
and an extremely important obstacle for increasing $\tilde{Z}T$.       $K_{pp}$ can be reduced, in principle, by
engineering the interfaces. Even without such improvement, according to the
geometry $K_{pp}\simeq 4K_p$ where $K_p$ is the bulk phonon
thermal conductance across the junction. In the relevant regime
$K_{pe}\simeq e^{-2}G_{in}\langle\omega\rangle^2 \gtrsim e^{-2}G_{in}E_g^2$. From Eq.
(\ref{ginap}) we then obtain
\be
K_{pe} \simeq \left(\frac{E_g}{k_BT}\right)^{2} 
e^{-\frac{E_g}{2k_BT}} E_g \zeta N_d\sum_\alpha D_\alpha L^{-1} .
\ee
We shall introduce yet another parameter to characterize the ratio
$K_e/K_p$ with $K_e$ being the electron thermal conductance
in a doped sample of the same geometry and  size as the junction, with
doping density $N_d$. The temperature dependence of the ratio
$K_e/K_p$ varies with the doping, structure, and
temperature. 
{The temperature dependence}
is assumed to be of the  form $K_e/K_p\propto T^\beta$ with
$\beta$ being a constant. A temperature-independent
prefactor parameter $\xi = K_eE_g^\beta/(K_pk_B^\beta T^\beta)$ is then
introduced. We further assume that the transport  properties in the
$n$ and $p$ regions are  similar (up to signs). The Wiedemann-Franz
law implies that $K_e=\eta (k_BT)^2 e^{-2}G_D$ where $G_D$ denotes the
electrical conductance in the doped sample and $\eta\simeq 2$ for
Boltzmann statistics. According to Ref.~\onlinecite{shockley},
$G_D\simeq 0.25G_{el}\exp [E_g/(k_BT)]$ when $N_d\simeq N_0$ and
$L/L_\alpha$ is small. Using Eq. (\ref{gelap}) one can write
$K_{pp}\simeq 4K_p\simeq 4\xi^{-1}K_e \simeq E_g^\beta
(k_BT)^{1-\beta}\xi^{-1}\eta N_0\sum_\alpha D_\alpha
L_\alpha^{-1}$. The figure of merit is then estimated as 
\bea
\tilde{Z}T &\simeq& \frac{K_{pe}}{K_{pp}}\simeq 
\frac{ \left(\frac{E_g}{k_BT}\right)^{2} 
e^{-\frac{E_g}{2k_BT}} E_g  \zeta N_0\sum_\alpha D_\alpha L^{-1}}{ E_g^\beta
  (k_BT)^{1-\beta}\xi^{-1} \eta
N_0\sum_\alpha D_\alpha L_\alpha^{-1}}\nn\\
&\simeq &  \left(\frac{E_g}{k_BT}\right)^{3-\beta} 
e^{-\frac{E_g}{2k_BT}} \frac{\zeta\xi}{\eta}\frac{L_\alpha}{L} .
\eea
If the temperature dependence comes mainly from the first two factors
then a high figure of merit can be obtained for $E_g/(2k_BT)=3-\beta$. 
For large $L_\alpha /L$ the three-terminal figure of merit can be larger than the two-terminal one. The power factor of the
device, $P= G_{in}S_p^2$, 
 can be  estimated similarly.

An especially appealing setup exploiting the
topological-insulator--ordinary-insulator--superlattices, can be 
built as follows (taking the superlattice  Bi$_2$Te$_3$/Si as an
example): On the front and back surfaces of each thin Bi$_2$Te$_3$
layer there are protected surface states with a gapless Dirac-cone like
spectrum.\cite{TIreview} Tunneling between the two surfaces opens a
gap in the spectrum of each surface band.\cite{burkov} In
superlattices these states form a pair of conduction and valence
minibands where the band gap can be controlled via the thickness of
the two types of layers [for details see Ref.~\onlinecite{burkov} and
footnote~\onlinecite{fotnot1}]. As the energy of the optical phonon in
Si is much larger than that in Bi$_2$Te$_3$, the optical phonons in Si
layers are well localized in these layers. So do the optical phonons in the
Bi$_2$Te$_3$ layers. Similarly, due to the significant mismatch of the
mechanical properties of the two types of layers, the acoustic phonons
also have  difficulty in being transmitted 
across the interfaces. The
phononic thermal conductivity along the growth direction is then
considerably reduced. It should be smaller than the values for both
materials in the bulk.\cite{lowkl} On the other hand, the phononic heat conductivity
within each thin layer of Si is large. At T=300~K, the former is
1.5~Wm$^{-1}$K$^{-1}$,\cite{bitekl} whereas the latter is about 
$10^2$~Wm$^{-1}$K$^{-1}$.\cite{sikl} When the electric
current is along the growth direction, the small phononic heat conductance 
along this direction greatly reduces the parasitic heat
conduction $K_{pp}$ and benefits the figure of merit. On the other
hand within each Si layer phonons are transferred efficiently from the
thermal terminal to the system, which is good for enhancing the output
power of the device.

We now estimate the figure of merit of the device using the example of 
the semiconductor made of  
the Bi$_2$Te$_3$/Si superlattice. We shall use the transport
parameters of bulk Bi$_2$Te$_3$ to do the estimation although the
superlattice should have better thermoelectric performance.\cite{bitesl}
First tune the superlattice structure to make
$E_g\lesssim\langle\omega\rangle$ with $\langle\omega\rangle$ being the energy of the optical
phonon in Si which is about 60~meV [see Ref. \onlinecite{LB}] (equivalent to
about 700~K). We will choose $E_g=600K$. In Bi$_2$Te$_3$  with $N_d\simeq
10^{20}$~cm$^{-3}$, one finds  from Ref.~\onlinecite{bitekl} 
that $\kappa_e\simeq \kappa_p$ at 300~K. This determines the parameter 
$\xi$ to be $2^\beta$. Using these parameters we calculate
and plot the figure of merit as a function of temperature in Fig.~\ref{fig2}
for a practically achievable value $L_\alpha/L=4$ and a modest
value of $\zeta=1/4$. In the calculation we ignored the
temperature dependence of $\zeta$ and $L_\alpha$. The results are
computed for three situations with $\beta = 1, 2$, and $2.5$. It is
seen that the figure of merit with the several underestimations made, 
can be larger than 1 around room
temperature for all the three situations, indicating potential
usefulness. Finally we note that the same strategy
and analysis can also be applied to the superlattice made of an
inverted-band gapless semiconductor and a normal semiconductor, where
the band gap can be tuned via the quantum well effect\cite{bhz} and
the normal semiconductor can be chosen to have the proper optical phonon
frequency and the high thermal conductivity to enable better
  thermoelectric performance.

\begin{figure}[htb]
  \centering \includegraphics[height=4.5cm]{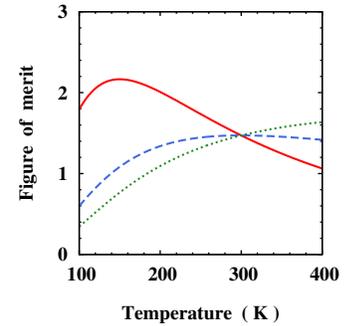}
  \caption{(Color online) The figure of merit $\tilde{Z}T$ of Bi$_2$Te$_3$/Si
    superlattices as a function of temperature for $\beta=1$ (solid
    curve), $2$ (dashed curve), $2.5$ (dotted curve). $E_g=600K$, 
    $\zeta = 1/4$, $\eta=2$, $L_\alpha/L=4$, and $\xi=2^{\beta}$ as
    $K_p=K_e$ at $T=300$~K.} \label{fig2}
\end{figure}

\section{Conclusions and discussions}

\label{SECCON}

We proposed and studied a  thermoelectric device using the phonon
bath as an example. The scheme is based on the ``three-terminal''
geometry of thermoelectric  
applications,\cite{3t0,3t1,3t2,3t,3t3} where inelastic processes play
a crucial role. It has been shown that a high thermoelectric 
figure of merit can be achieved in this geometry in several
nanosystems,\cite{3t} where only one microscopic energy channel
in which the relevant electronic energy is fixed, is available. In this paper we
derived the figure of merit for the multiple energy channel case. We find that,
when only the inelastic processes are considered, the figure of merit
is the ratio of the square of the mean value of {\em energy
  difference} between final and initial states to its variance with
the average weighed by the conductance of each microscopic process. A
small variance in the energy change is then favorable for a  high
figure of merit. To achieve such a good energy selection, one can use
either the electronic band gap, $E_g\gg k_BT$, for electrons or the
narrowness of the phonon band, $\gamma\ll \ave{\ome}$, for
phonons. The realistic figure of merit including other
processes, Eq.~(\ref{zt3t}), is also discussed. It is found that a
strong carrier-boson coupling as well as the dominance of the
  inelastic transport  and a small purely phononic thermal conductivity {\em between the phonon terminal and the electronic ones}, are necessary for a high figure of
merit. The suppression of the elastic transport can be achieved
with a semiconductor junction,\cite{shockley} while
  the coupling is strong when the bosons are, e.g., phonons,
  electron-hole pair excitations, etc.. Thanks to
those, the proposed three-terminal device can have a figure of merit higher than that of 
the usual two-terminal device made of the {\em same} material. 

In comparison with the existing literature, M-S suggested a narrow
electronic band for elastic two-terminal transport to achieve high
values of $ZT$. When this scheme is  generalized to  inelastic
processes where the initial $E_i$ and final $E_f$ energies are
different,  high values of   $ZT$}   are possible when the distribution of
the average energy $\ov{E}=(E_i+E_f)/2$   (measured from the common chemical potential)  is narrow in the two-terminal
geometry.  In the three-terminal situation  a narrow distribution of 
$\ome=E_f-E_i$ plays a crucial role. 
The latter can also be
achieved by controlling the initial and final electronic states by a barrier
higher enough than $T$, or in a small system with 
a few 
initial and final states with fixed $\ome=E_f-E_i$, as in
Refs.~\onlinecite{3t,3t2,3t3}, or, for example, via the narrow bandwidth of optical
phonons. Finally, we note the analogy of the suggested configuration,
which may convert thermal to electrical energy, and a photovoltaic device.

\section*{Acknowledgments} 
OEW acknowledges the support of the Albert Einstein Minerva Center for
Theoretical Physics, Weizmann Institute of Science. This work was
supported by the BMBF within the DIP program, BSF, by ISF, and by its
Converging Technologies Program.


\begin{thebibliography}{99}
\bibitem{note11} Here we have used certain simplifications. The
  consideration only holds for the central part of the intrinsic
  region, while the situation near the boundaries is more
  complicated. However, this is not supposed to change the results
  considerably.\cite{closed}
\bibitem{fotnot1} In Ref.~\onlinecite{burkov} it is shown that the gap
  is $E_g=2|\Delta_S-\Delta_D|$ where $\Delta_S$ and $\Delta_D$ are  the tunneling rates
  through the topological and the ordinary insulator layer, respectively.  They (and the
 band gap) can be tuned in a large range via the thickness of these  layers.
\end{thebibliography}
\end{document}